\documentclass[preprint,12pt]{elsarticle}

\usepackage{amssymb}
\usepackage{amsmath}
\usepackage{url}
\usepackage{float}
\usepackage{placeins}

\journal{Information Sciences }

\newif\ifblinded
\blindedfalse

\newif\ifarxiv
\arxivtrue

\begin{document}

\begin{frontmatter}



\title{Good-Enough LLM Obfuscation (GELO)}


\author[inst1]{Anatoly Belikov} 
\ead{abelikov@singularitynet.io}

\author[inst2]{Ilya Fedotov} 
\ead{ilya@singularitycompute.com}

\affiliation[inst1]{organization={SingularityNET Foundation},
    addressline={Baarerstrasse 141},
    city={Zug},
    postcode={6300},
    country={Switzerland}}

\affiliation[inst2]{organization={Singularity Compute},}

\begin{abstract}
Large Language Models (LLMs) are increasingly served on shared accelerators where an adversary with read
access to device memory can observe KV caches and hidden states, threatening prompt privacy for
open-source models. Cryptographic protections such as MPC and FHE offer strong guarantees but remain
one to two orders of magnitude too slow for interactive inference, while static obfuscation schemes break
under multi-run statistical attacks once the model is known. We present GELO (Good-Enough LLM
Obfuscation), a lightweight protocol for privacy-preserving inference that limits \emph{information leakage}
from untrusted accelerator observations by hiding hidden states with fresh, per-batch invertible mixing. For
each offloaded projection, the TEE samples a random matrix $A$, forms
$U = A H$, offloads $U$ and weights $W$ to the accelerator, and then applies $A^{-1}$ on return, so that
$A^{-1}((A H) W) = H W$ and outputs are unchanged.  We analyze information leakage and introduce two practical defenses: (i) non-orthogonal mixing to mask Gram matrices, and (ii) orthogonal
mixing augmented with a small fraction of high-energy ``shield'' vectors that pollute higher-order
statistics. On Llama-2 7B, GELO preserves float32 outputs exactly, closely matches low-precision
baselines, and shows about $20$--$30\%$ compute-side overhead in a controlled offload microbenchmark;
an unoptimized remote prototype is dominated by transport overhead, motivating deeper serving-engine
integration.  GELO resists ICA/BSS and anchor-assisted attacks; a 60M-parameter transformer-based unmixing attack also fails under strong mixing and shielding.

\end{abstract}





\end{frontmatter}

\section*{CRediT authorship contribution statement}
\noindent Anatoly Belikov: Conceptualization, Methodology, Software, Formal analysis, Investigation, Validation, Writing---original draft.
Ilya Fedotov: Supervision, Funding acquisition, Resources, Writing---review \& editing.



\section{Introduction}
\label{sec1}
Large Language Models (LLMs) are increasingly deployed on shared cloud GPUs. In this setting, an adversary with read access to device
memory can exploit vulnerabilities such as KV-cache leakage to reconstruct confidential prompts, infer user data, or partially
reverse-engineer model behavior. This creates a tension between the scalability benefits of cloud inference and strict privacy
requirements on user inputs.

Existing approaches lie on two extremes. Cryptographic methods such as Fully Homomorphic Encryption (FHE) and Multi-Party Computation (MPC)
provide strong, provable guarantees, but they typically incur over $100\times$ latency overhead and remain impractical for interactive LLM
services. At the other end, lightweight obfuscation schemes based on static permutations of weights or activations are fast, but fragile:
once the underlying model is known, they can be broken by multi-run statistical attacks.

Our primary target deployment is a mixed cluster with a small number of confidential GPUs (e.g., H200 with TEE support)~\cite{nvidia_confidential_computing} and a larger pool
of non-confidential accelerators (e.g., L40S). Privacy requirements disallow running plaintext hidden states on the L40S, yet relying
solely on H200s for full inference would cap cluster throughput. We therefore seek a protocol that keeps sensitive data inside TEEs while
still letting the L40S execute most of the heavy linear algebra.

To this end, we introduce the Good Enough LLM Obfuscation (GELO) protocol. GELO is a hybrid design: it executes most transformer
operations inside a Trusted Execution Environment (TEE), while offloading only the most expensive matrix multiplications in self-attention,
namely the Query (Q), Key (K), and Value (V) projections, to an untrusted accelerator. For each batch, GELO applies a fresh secret
invertible linear transform $A$ to hidden states $H$ before offload. The accelerator computes projections on the mixed data
$U = A H$, and the TEE applies $A^{-1}$ to recover the exact result. Because $A$ is never reused, the attacker faces a single-batch Blind
Source Separation (BSS) problem. Our controlled microbenchmarks show that the added mixing/unmixing computation is modest, while our remote
prototype shows that end-to-end latency is dominated by systems integration and transport overheads.

Our key contributions are:
\begin{itemize}
	\item \textbf{The GELO Protocol:} We introduce and formalize the Good Enough LLM Obfuscation (GELO) algorithm, a lightweight protocol for
offloading LLM projection computations to an untrusted accelerator without revealing the underlying hidden states.
	\item \textbf{Leakage and identifiability analysis:} We identify key leakage channels (e.g., Gram-matrix invariants under orthogonal
mixing) and provide an identifiability-based security argument: per-batch non-identifiability up to an unknown invertible transform, and no
cross-batch information gain under fresh, independent mixing.
	\item \textbf{Attack evaluation and efficiency:} We empirically evaluate anchor-based, BSS/ICA, and transformer-based learned recovery attacks under this threat
model, and quantify both compute-side overhead and unoptimized end-to-end prototype overhead, showing that GELO's core operations are
lightweight while production latency depends on serving-runtime integration.
\end{itemize}

The remainder of this paper is structured as follows. Section~\ref{sec:background} (Background and Related Work) reviews related work in
cryptographic, obfuscation-based, and TEE-assisted private inference. Section~\ref{sec:gelo} (The GELO Protocol: Method and
Implementation) provides a detailed description of our protocol. Section~\ref{sec:experiments} (Experiments) presents an analysis of
performance and security properties. We also analyze what information about hidden states is leaked by accelerator-visible observables under
GELO, and characterize the conditions under which an attacker could invert the mixing. Section~\ref{sec:security} (Security Analysis and
Identifiability) concludes and outlines directions
for future work.

\section{Background and Related Work}
\label{sec:background}

The challenge of securing LLM inference has been approached from several directions, primarily falling into three categories:
cryptographic, obfuscation-based, and TEE-assisted methods. Each offers a different balance of security, performance, and practicality.

\subsection{Cryptographic Approaches}

Cryptographic techniques like Multi-Party Computation (MPC) and Fully Homomorphic Encryption (FHE) represent the gold standard for
security, allowing computation on data without ever exposing it in plaintext. In an MPC-based system, secret shares of the inputs and model
weights are distributed among multiple non-colluding parties, who collaboratively compute the result. FHE, in theory, allows a single
untrusted server to perform calculations directly on encrypted data.

While these methods provide powerful security guarantees, their practical application to large-scale transformers is severely hampered by
performance overhead. The non-linear operations ubiquitous in transformers, such as Softmax and GELU activations, are notoriously expensive
to compute in both MPC and FHE frameworks. For instance, Fission's MPC approach~\cite{ugurbil2025fission} scales poorly for attention
mechanisms due to the high communication and computation cost of its secure non-linear function protocols, leading to latencies that are
orders of magnitude slower than plaintext inference and unsuitable for interactive applications. The fundamental security assumption of MPC
is that the nodes are run by separate, independent, and non-colluding entities. This makes the trust assumption difficult to satisfy in
many practical deployments.

\subsection{Hybrid and Obfuscation-based Approaches}

To avoid the extreme overhead of cryptography, obfuscation-based methods apply transformations to the model or data to make them
unintelligible to an observer. These techniques range from permutation-based protocols to hybrid TEE-based approaches, each with different
trust models and security guarantees.

A prominent protocol is the Secure Transformer Inference Protocol (STIP)~\cite{yuan2023stip}. 
STIP is a three-party protocol involving a User, a Model Owner, and an untrusted Executor who performs the live computation. 
The security hinges on a static secret permutation matrix ($\pi$) generated by the User. 
The User provides $\pi$ to the Model Owner, who transforms the private model weights ($\theta$) 
into obfuscated weights ($\theta' = f(\theta,\pi)$) and sends them to the Executor. 
The User then obfuscates their prompt ($x' = x\pi$) and sends it to the Executor. 
The insight is that the permutations are designed to cancel out internally (e.g., $Q = (x\pi)(\pi^\top W) = x W$), 
ensuring correctness. However, STIP's architecture has two severe limitations: 
its rigid three-party trust model is impractical for many cloud scenarios, and more critically, 
STIP's security collapses for open-source models. If the original weights ($W$) are public, 
an adversary can solve the equation $W' = \pi^\top W$ to recover the user's secret permutation $\pi$.

PermLLM~\cite{zheng2024permllm} represents a different, hybrid-MPC approach. It also uses a three-party model, but with different roles: a
User $P_1$, a Model Provider who also acts as the Hoster $P_0$, and a third party $P_2$ that assists only in an offline preparation phase.
The live inference is a two-party cryptographic computation between the User and the Hoster. Its mechanism is not a simple static
permutation; instead, it uses additive secret sharing (an MPC technique) for all linear layers and reserves a secure random permutation for
non-linear functions (like Softmax). This permutation is dynamic and allows the User to compute the non-linear function on shuffled
plaintext data, avoiding the primary MPC bottleneck. While this design is “magnitudes faster” than pure MPC and is secure for open-source
models (user privacy is protected by secret sharing, not key obfuscation), it still retains the high network communication overhead and
latency inherent to cryptographic protocols for every linear layer.

STIP's rigid model and its critical vulnerability to open-source models make it incompatible with common deployments. 
PermLLM, while secure, retains the high latency of network-bound cryptographic protocols.
Our protocol, GELO, is designed to overcome these specific limitations. 
Like PermLLM, it is secure for open-source models. 
However, it achieves dramatically higher performance by rejecting high-overhead MPC entirely.
Instead, GELO leverages a hybrid TEE–accelerator model. 
By using a dynamic, per-batch secret linear transformation (an invertible matrix $A$) 
generated within the TEE, GELO can safely offload the vast majority of computation (the expensive matrix multiplies) 
to an untrusted accelerator. This approach, rooted in the computational hardness of Blind Source Separation (BSS), 
thwarts the statistical attacks that defeat static permutation schemes while avoiding the network latency of MPC, 
as will be detailed in Section~\ref{sec:gelo}.

\subsection{TEE-Based Approaches}

Trusted Execution Environments (TEEs), such as Intel SGX and AMD SEV, offer a compelling middle ground by providing hardware-isolated
enclaves where code and data are protected from the host system. This enables a practical two-party model (user and cloud provider).
However, TEEs typically have limited memory and cannot match the raw performance of high-end GPUs, necessitating hybrid approaches that
offload computation.

KV-Shield~\cite{yang2024a} is one such method, designed to protect on-device LLM inference from KV cache leakage. It uses a hybrid TEE–GPU
approach where a secret random permutation matrix $R$ is generated and stored in the TEE. This matrix $R$ is used to permute the
attention's linear weights ($W_P = W R$). Consequently, the computations on the untrusted GPU produce a permuted KV cache $K_P, V_P$,
theoretically protecting the original data from being leaked. The TEE applies an inverse permutation to the final result to ensure
correctness.

However, this design shares the exact same fundamental vulnerability as STIP. Its security hinges on the secrecy of $R$. In an open-source
model setting, an attacker knows the original weights $W$ and can observe the permuted $W_P$ as it's loaded onto the untrusted GPU for
computation. This allows them to solve the equation $W_P = W R$ to recover the secret $R$, completely compromising the protocol's privacy
guarantees.

The vulnerability exploited by such attacks is not merely theoretical; the LeftoverLocals attack~\cite{woot2024leftoverlocals} demonstrated
a practical method for intercepting the KV cache from GPU local memory to reconstruct LLM responses. This has spurred the development of
several defenses. KV-Cloak~\cite{luo2025shadowcache}, for instance, was proposed as another lightweight defense specifically targeting this
KV-cache privacy risk. Both protocols build on a legacy of hybrid TEE–GPU systems, such as ShadowNet~\cite{sun2023shadownet}, which was
originally designed for CNNs. This earlier work established the core strategy of partitioning a model, offloading computationally heavy
linear layers to an untrusted GPU while processing sensitive non-linear activations inside the TEE. However, as demonstrated by the
analysis of TEESlice~\cite{zhang2022teeslice}, this partitioning strategy is inherently vulnerable when the attacker has access to public
model information, a condition that holds for open-source LLMs. Beyond these specific algorithmic flaws, all TEE–GPU hybrid models face
severe practical challenges, including significant communication overhead from encrypting data transfers over PCIe and potential security
gaps in the accelerators themselves, such as unencrypted HBM memory on some confidential GPU models.

\section{The GELO Protocol: Method and Implementation}
\label{sec:gelo}

We propose a secure inference protocol that protects user privacy by executing most LLM operations within a Trusted Execution Environment
(TEE), while strategically offloading the most intensive computations to an untrusted accelerator. The protocol’s security is rooted in a
novel application of dynamic orthogonal rotations, which obfuscate intermediate data without altering the model's final output.

\subsection{System Architecture and Threat Model}

\begin{figure}[!htbp]
	\centering
	\includegraphics[width=0.95\linewidth]{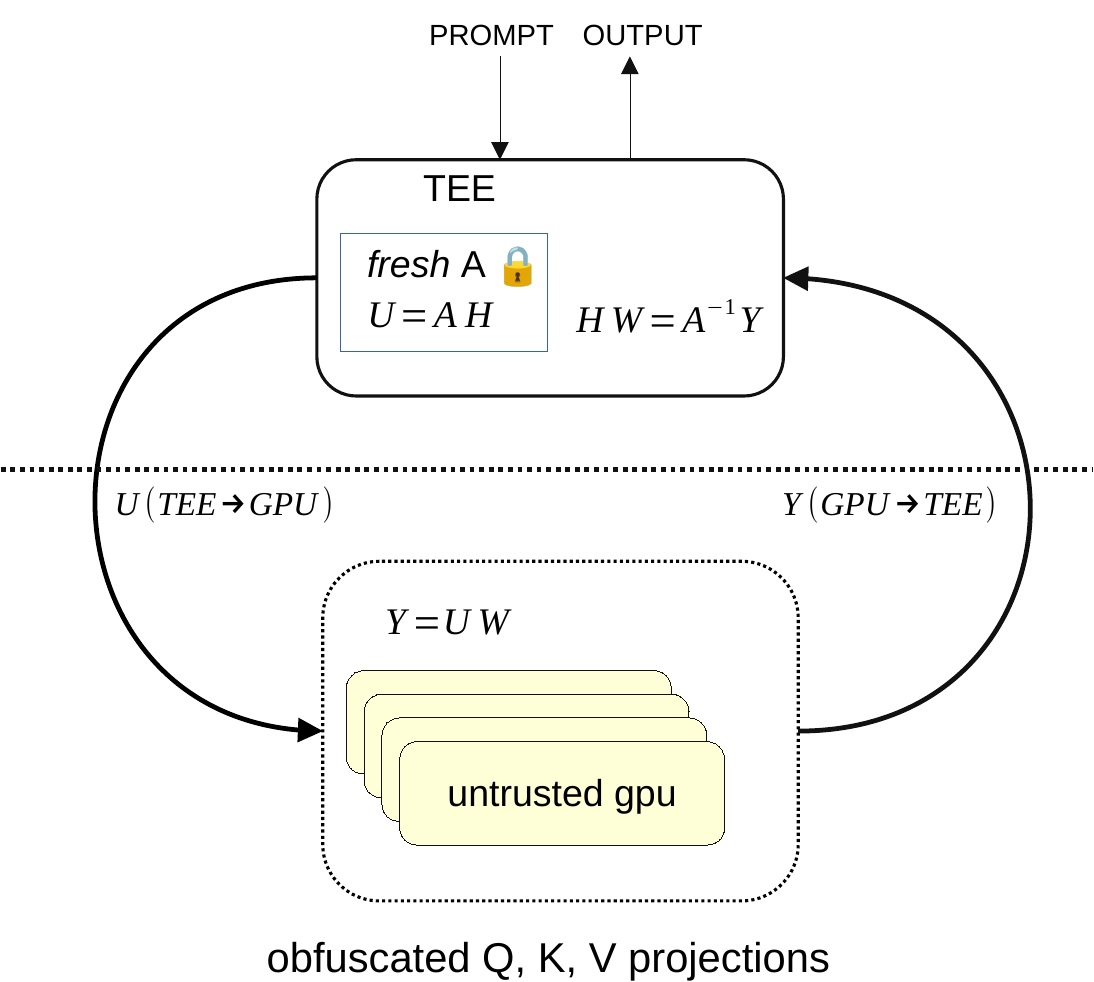}
\caption{GELO protocol overview. The diagram illustrates the $Q/K/V$ projections; the same procedure is also applied to the output projection ($O$).}
\label{fig:arch}
\end{figure}

Our system architecture consists of two distinct components (Figure~\ref{fig:arch}):
\begin{itemize}
	\item \textbf{Trusted TEE device (confidential GPU).} A TEE-enabled accelerator (e.g., H100/H200-class confidential GPU) where protocol
secrets and sensitive intermediate activations are stored and processed. The TEE is responsible for generating and managing all
cryptographic secrets.
	\item \textbf{Untrusted Accelerator.} A high-performance device (e.g., GPU) treated as completely untrusted and used only to execute
specific, computationally demanding matrix multiplications offloaded from the TEE. It knows the model architecture and weights (e.g.,
open-source LLMs) and all protocol details except ephemeral secrets.
\end{itemize}

We adopt an honest-but-curious threat model, standard in cloud settings. The adversary (e.g., a malicious cloud provider) has full,
real-time read access to the accelerator's memory (VRAM) and can observe all data transferred to and from it. The adversary's primary goal
is to compromise user privacy by reconstructing the input prompt from these observations. We assume the TEE provides strong hardware-level
confidentiality and integrity guarantees, protecting the secrets and computations within it. Side channels originating inside the TEE
(fine-grained cache timing, microarchitectural leakage, physical attacks) and availability attacks are out of scope. We also treat
accelerator-side leakage beyond raw memory reads (e.g., timing/cache side channels) as out of scope for this work~\cite{nvidia_confidential_computing,naghibijouybari2018rendered,woot2024leftoverlocals}.

Formally, for a protected batch $t$, let $H_t \in \mathbb{R}^{n_t \times d}$ denote the plaintext hidden-state rows held inside the TEE and
let $A_t \in \mathbb{R}^{n_t \times n_t}$ be a fresh secret invertible mixing matrix sampled independently for that batch. For an offloaded
projection with public weight matrix $W \in \mathbb{R}^{d \times p}$, the accelerator observes
\[
\mathcal{O}_t = \{U_t = A_t H_t,\; W,\; Y_t = U_t W,\; n_t,\; d,\; p,\; \Pi\},
\]
where $\Pi$ denotes public protocol details, model architecture, tokenizer, and offload schedule. The adversary does not observe $A_t$,
$A_t^{-1}$, plaintext $H_t$, shield rows before mixing, or any TEE-resident intermediate states. Across batches, $A_t$ is sampled freshly, so
observations $\mathcal{O}_1,\ldots,\mathcal{O}_T$ do not share a fixed mixing coordinate system.

The privacy-relevant attack goal is recovery of sensitive prompt information, such as names, dates, passwords, or other confidential strings.
We distinguish this from weaker exploratory goals used to probe leakage. In increasing order of strength, we consider:
	\begin{description}
		\item[Text-presence detection.] Decide whether a source text or document contributed rows to the mixed batch. This is an aggregate
		leakage test and does not by itself reconstruct token embeddings or prompt content.
		\item[Stream association.] In a streaming serving implementation, the adversary may observe individual rows $u$, where each $u$ is a row of
		some unknown mixed batch $U_t$, rather than complete matrices. The attacker must therefore collect observed rows and associate them with the corresponding mix $U_t$ before attempting unmixing.
		\item[Hidden-state reconstruction.] Process a reconstructed mixed batch $U_t$ and output $\hat{H}_t$ close to the true rows of $H_t$ up to
		permutation and matching ambiguity.
		\item[Sensitive-content recovery.] Use reconstructed hidden states to recover sensitive prompt information such as names, dates, credentials, or private facts.
	\end{description}
		An end-to-end learned attack against a real serving trace would likely need to combine several stages. The adversary would first collect
streamed accelerator-visible rows $u$, use an association model to decide which rows plausibly belong to the same underlying mixed batch
$U_t$, use a separate model to estimate hidden rows $H_i$ from that reconstructed $U_t$, and only then map recovered rows to source texts or
tokens. Our transformer-based text-presence experiment in Section~\ref{sec:learned-attack} targets text presence as a conservative leakage
probe on already constructed mixtures, while the token-level unmixing experiment targets the harder hidden-state reconstruction step under a
favorable attacker setting where the relevant mixed set is already available. Since unmixing fails under strong fresh mixing and shielding even
under these favorable conditions in our experiments, we do not perform a full end-to-end attack.

\subsection{Core Algorithm}
\label{sec:core-algorithm}

The core of the GELO protocol is a sequence of linear-algebraic operations designed to offload linear projections while keeping the
sensitive hidden-state matrix $H$ confidential. The protocol is executed for each attention block where offloading is desired.

Before the protocol begins, the TEE applies batching-time defenses to harden against known-plaintext and data-amplification attacks:
\begin{itemize}
	\item \textbf{Cross-user mixing.} Construct batches by aggregating requests from multiple independent users to reduce per-user signal
correlation.
	\item \textbf{Flooding detection and disruption.} Maintain token-frequency statistics; if they significantly diverge from a baseline
(e.g., repeated known tokens), inject random tokens to disrupt the pattern.
	\item \textbf{Sensitive-layer exclusion.} Do not apply GELO to the first few layers nor the final layer; compute them entirely inside the
TEE.
\end{itemize}

The offloading protocol for a single projection proceeds as follows. Let $H \in \mathbb{R}^{n \times d}$ be the batched hidden states (with
$n$ tokens, model dimension $d$), and let $W \in \mathbb{R}^{d \times p}$ be the projection matrix.

\begin{description}
	\item[1. Trusted side (TEE).]
	\begin{enumerate}
		\item \textbf{Define inputs.} $H \in \mathbb{R}^{n \times d}$, $W \in \mathbb{R}^{d \times p}$.
		\item \textbf{Generate secret matrix.} Sample a fresh, random invertible matrix $A \in \mathbb{R}^{n \times n}$ for this batch (never
reused across batches).
		\item \textbf{Obfuscate data (mixing).} Compute
		\begin{equation}
			U = A H \, .
		\end{equation}
		\item \textbf{Offload computation.} Send $U$ (and $W$ if not already resident) to the accelerator.
	\end{enumerate}
	
	\item[2. Untrusted accelerator.]
	\begin{enumerate}
		\item \textbf{Perform projection (general matrix multiplication; GEMM).} Compute
		\begin{equation}
			Y = U W \, .
		\end{equation}
		From the adversary's perspective, both $U$ and $W$ are visible in VRAM.
		\item \textbf{Return result.} Send $Y \in \mathbb{R}^{n \times p}$ back to the TEE.
	\end{enumerate}
	
	\item[3. Trusted side (TEE).]
	\begin{enumerate}
		\item \textbf{De-obfuscate (un-mixing).} Recover the true projection via
		\begin{equation}
			Q = A^{-1} Y \, .
		\end{equation}
		If $A$ is orthogonal, $A^{-1} = A^\top$.
	\end{enumerate}
\end{description}

Correctness follows immediately:
\begin{equation}
	Q \;=\; A^{-1} Y \;=\; A^{-1} (U W) \;=\; A^{-1} (A H W) \;=\; (A^{-1} A) H W \;=\; I H W \;=\; H W \, .
\end{equation}

\subsubsection{Mitigating Gram Matrix Information Leaks}
\label{sec:gram-leaks}

If $A$ is orthogonal (desirable for performance, as $A^{-1} = A^\top$), an adversary observing $U = A H$ can compute:
\begin{align}
	U^\top U &= (A H)^\top (A H) \;=\; H^\top A^\top A H \;=\; H^\top H \, , \\
	U U^\top &= (A H)(A H)^\top \;=\; A (H H^\top) A^\top \, .
\end{align}
Thus:
\begin{itemize}
	\item \textbf{Covariance leak.} $U^\top U = H^\top H$ perfectly reveals the $d \times d$ covariance of hidden states.
	\item \textbf{Similarity-spectrum leak.} $U U^\top$ is similar to $H H^\top$ and shares its eigenvalues, leaking the spectrum of
token-token similarities.
\end{itemize}
While this does not directly reconstruct $H$, it may be unacceptable. We consider two mitigations.

\paragraph{Mitigation 1: Use a non-orthogonal $A$.}
Choose $A$ to be a general invertible matrix, so that $A^\top A \neq I$ and the Gram matrices are masked.
\begin{itemize}
	\item \textbf{Computational cost.} The TEE must compute $A^{-1}$ per batch, an $\mathcal{O}(n^3)$ operation.
	\item \textbf{Numerical stability.} Orthogonal $A$ has condition number $\kappa(A)=1$. A general $A$ can have $\kappa(A)\gg 1$, amplifying
accelerator-side numerical errors (e.g., BF16/FP16) in $Q = A^{-1} Y$. The TEE should generate and verify well-conditioned $A$ (e.g.,
enforce $\kappa(A)<100$).
\end{itemize}

\paragraph{Mitigation 2: Random vector padding (shielding).}
Retain a fast, stable orthogonal $A \in \mathbb{{R}}^{n \times n}$ but pad the batch with $k$ random “shield” vectors:
\begin{align}
	&H_{\text{data}} \in \mathbb{R}^{(n-k)\times d}, \quad S \in \mathbb{R}^{k \times d}, \\
	&H_{\text{full}} = \begin{bmatrix} H_{\text{data}} \\ S \end{bmatrix} \in \mathbb{R}^{n \times d}, \quad U = A H_{\text{full}} \, .
\end{align}
Then the accelerator observes
\begin{equation}
	U^\top U \;=\; H_{\text{full}}^\top H_{\text{full}} \;=\; H_{\text{data}}^\top H_{\text{data}} + S^\top S \, ,
\end{equation}
i.e., only the sum of the real covariance and the shield covariance. Since $S$ is secret and fresh, the attacker cannot isolate
$H_{\text{data}}^\top H_{\text{data}}$. The same “pollution” masks the token-similarity spectrum.

The trade-off is reduced data throughput: inserting $k$ shield vectors displaces $k$ user tokens for the same $\mathcal{O}(n^2 d)$ cost.
Empirically, adding a small number of shield vectors ($k \approx 5\%$ of $n$) and scaling them to 4–10× the average row norm of
$H_{\text{data}}$ thwarts ICA-based attacks with negligible throughput penalty.

\section{Experiments}
\label{sec:experiments}

\subsection{Functional Equality}

To validate the mathematical correctness of the GELO protocol, we conducted an end-to-end functional equality test. We compared the output
logits of a baseline Llama~2 7B model~\cite{llama2} against an identical model where GELO obfuscation was applied to all $Q$, $K$, and $V$
projections, as well as the attention output projection ($O$), in the attention layers. Both models were evaluated on a 1000-sample subset
of the OpenWebText2 dataset.

We measured correctness using three metrics: (i) the equality rate of the top-1 predicted token, (ii) the mean squared error (MSE) between
the final logit vectors, and (iii) the mean per-token $L_2$ distance between the logit vectors. Results are shown in
Table~\ref{tab:functional-equality}.

\begin{table}[t]
	\centering
	\caption{Functional equality between baseline and GELO-obfuscated inference across precisions.}
	\label{tab:functional-equality}
	\begin{tabular}{lccc}
		\hline
		Precision & Top-1 Equality & Logit MSE & Mean per-token $L_2$ \\
		\hline
		float32  & 1.000000 & $9.320817 \times 10^{-11}$ & 0.000902 \\
		bfloat16 & 0.988045 & $1.813321 \times 10^{-3}$  & 6.179683 \\
		float16  & 0.998450 & $4.312208 \times 10^{-5}$  & 0.793820 \\
		\hline
	\end{tabular}
\end{table}

In float32 precision, the protocol achieves perfect equality, with a 100\% match in top-1 tokens and near-zero MSE, confirming
\begin{equation}
	A^{-1}\!\big((A H) W\big) \;=\; H W \, .
\end{equation}
In practical low-precision formats (bfloat16 and float16), GELO introduces no meaningful numerical error: the top-1 token equality remains
above 98.8\%, indicating that the minimal precision loss from the extra mix/un-mix operations does not degrade the model's generative
output in any practical sense.

\subsection{Performance and Latency Analysis}

\paragraph{Experimental setup}
End-to-end serving overhead depends heavily on inference-engine details such as KV-cache management, batching/scheduling policies, kernel
fusion, memory layout, asynchronous execution, and transport implementation (e.g., in vLLM). Integrating GELO into such an engine would
require substantial software engineering effort beyond the scope of this study. We therefore separate two questions: (i) the algorithmic and
compute-side overhead of GELO's mix/unmix operations, measured in a controlled microbenchmark, and (ii) the end-to-end overhead of an
unoptimized remote prototype, which exposes the systems bottlenecks that a production implementation would need to address.

Our synthetic microbenchmark uses a same-machine logical split to model the trusted/untrusted components. We obfuscate and transmit random
batches between two processes running on different GPUs, rather than running end-to-end LLM inference. This setup allows us to measure the
full stack overheads relevant to GELO-style offload, including $A$-generation, mixing/unmixing, and communication (IPC) latency.
The experimental code is available at \url{https://github.com/noskill/gelo}.

\paragraph{Compute-side overhead vs. batch size}
We first measured the latency overhead of GELO versus an insecure baseline (direct offload without obfuscation) across batch sizes $n$ in the
controlled two-process microbenchmark (Table~\ref{tab:overhead-batch}). This experiment includes $A$-generation, mixing/unmixing, GEMM, and
same-machine communication, but should not be interpreted as a full end-to-end serving benchmark.

\begin{table}[t]
	\centering
	\caption{Latency and overhead versus batch size.}
	\label{tab:overhead-batch}
	\begin{tabular}{rccc}
		\hline
		Batch size $n$ & Overhead (\%) & GELO total (ms) & Baseline (ms) \\
		\hline
		64  & 28.9 &   4.49  &   3.48 \\
		128  & 29.2 &   5.79  &   4.48 \\
		256  & 19.9 &   8.93  &   7.44 \\
		512  & 20.1 &  16.11  &  13.41 \\
		1024  & 26.2 &  30.36  &  24.06 \\
		2048  & 32.8 &  60.31  &  45.41 \\
		4096  & 32.8 & 218.75  & 164.74 \\
		8192  & 50.1 & 537.85  & 358.41 \\
		\hline
	\end{tabular}
\end{table}

The results reveal a U-shaped overhead curve:
\begin{itemize}
	\item For small batches ($n < 128$), overhead is high ($\sim 29\%$) because GELO-specific costs (A-generation, mixing) are large relative
to the very fast main GEMM.
	\item At $n \in \{256, 512\}$, overhead is minimized ($\sim 20\%$). Here, the $\mathcal{O}(n d^2)$ GEMM dominates, making GELO’s costs a
smaller fraction of total time.
	\item For large batches ($n > 2048$), overhead rises as the $\mathcal{O}(n^3)$ cost of generating the $n \times n$ orthogonal matrix $A$
becomes the bottleneck.
\end{itemize}

\paragraph{Latency breakdown ($n=512$)}
To understand the overhead sources, we profiled a single run at $n=512$ (Table~\ref{tab:latency-breakdown}).

\begin{table}[t]
	\centering
	\caption{Latency breakdown at $n=512$.}
	\label{tab:latency-breakdown}
	\begin{tabular}{lccc}
		\hline
		Step & GELO (ms) & \% of total & Baseline (ms) \\
		\hline
		A-gen (QR)           &  2.322 & 13.3\% & --    \\
		Mix ($A \cdot H$)    &  0.199 &  1.1\% & 0.000 \\
		GEMM ($U \cdot W$)   &  0.441 &  2.5\% & 0.443 \\
		Un-mix ($A^{-1}\!\cdot Y$) &  0.272 &  1.6\% & 0.000 \\
		Copy (socket+I/O)    & 14.186 & 81.4\% & 14.123 \\
		\hline
		Total                & 17.420 & 100.0\% & 14.566 \\
		\hline
	\end{tabular}
\end{table}

In this controlled setting, total overhead at $n=512$ is 19.6\%. Two key insights emerge:
\begin{itemize}
	\item \textbf{Modest compute cost.} The computational overhead of GELO is A-gen + Mix + Un-mix = 2.793 ms, representing the true cost of
security, which is modest.
	\item \textbf{Communication bottleneck.} The majority of time ($\sim 81\%$) in both GELO and the baseline is spent on Copy (socket+I/O),
indicating the experiment is bottlenecked by inter-process communication rather than GELO’s computations.
\end{itemize}

In summary, the controlled microbenchmark shows that GELO's core computation can be added with about 20\% overhead near the best-performing
batch sizes, while successfully offloading the main GEMM. This number should be read as a compute-side estimate for the protocol mechanics,
not as an end-to-end serving overhead claim.

\paragraph{End-to-end remote prototype}
We also implemented a prefill-only asynchronous remote prototype on a 4xRTX 3090 machine, where GPU 0 runs the trusted model process and GPUs
1--3 act as untrusted workers for obfuscated Q/K/V projections. With CodeLlama-7B, batch size 2, sequence length 256, and request concurrency
6, the local baseline reached 3679 tok/s, while the three-worker async remote prototype reached 669 tok/s, corresponding to 450\% overhead.
The run transferred 4.0 GiB of obfuscated input activations and 12.2 GiB of returned Q/K/V activations over 1024 RPC calls; remote GEMM
accounted for only 1.76\% of accumulated RPC time.

This prototype result should be interpreted as a systems baseline rather than an optimized GELO deployment. The overhead is dominated by
Python/socket serialization, RPC scheduling, and activation transfer, not by GELO's mixing or projection computation. A production
implementation would need deeper serving-runtime integration: asynchronous request scheduling, fused serialization/transport, persistent GPU
buffers, peer-to-peer or RDMA-style transfers where available, and overlap between communication and trusted-side computation. We therefore
view the prototype as evidence that the core protocol is computationally lightweight, while the end-to-end overhead is primarily an engineering
optimization problem outside the scope of this paper.

\subsection{Deobfuscation Security Analysis: Hidden State Statistics}
\label{sec:deobfuscation-stats}

The security of GELO rests on the infeasibility of solving the Blind Source Separation (BSS) problem $U = A H$ from a single observation
$U$. However, BSS algorithms (e.g., ICA, Dictionary Learning) can be effective when the sources $H$ exhibit exploitable statistical
structure or strong priors. We therefore analyze hidden-state statistics to identify properties an adversary might leverage.

Our dataset comprises 10 million embedding vectors extracted from the 10th transformer layer of a Llama~2 7B model running on the
OpenWebText2 dataset.

\subsubsection{Known-Plaintext Vulnerability via Token Repetition}

A classic attack against obfuscation is known-plaintext: an adversary injects or exploits repeated, known inputs
and maps them to their obfuscated outputs. We first examined our 10M-embedding dataset for exact duplicates. The results are shown in
Table~\ref{tab:duplicates}.

\begin{table}[t]
	\centering
	\caption{Duplicate analysis over 10{,}000{,}165 embeddings.}
	\label{tab:duplicates}
	\begin{tabular}{lrr}
		\hline
		Metric & Value & Percentage \\
		\hline
		Total Embeddings   & 10{,}000{,}165 & 100\% \\
		Unique Embeddings  & 8{,}359{,}390  & 83.6\% \\
		Duplicate Embeddings & 1{,}640{,}775 & 16.4\% \\
		\hline
	\end{tabular}
\end{table}

While a 16\% duplication rate appears high, frequency analysis shows these duplicates are dominated by special tokens (BOS/EOS). Removing
special tokens reduces the collision rate to 0.148\%.
Table~\ref{tab:top5hashes} lists most frequent hashes after we remove \textit{bos} and \textit{eos} tokens.

\begin{table}[t]
	\centering
	\caption{Top 5 most frequent embedding hashes.}
	\label{tab:top5hashes}
	\begin{tabular}{rllll}
		\hline
		\# & Hash & Count & Token IDs & Tokens \\
		\hline
		1 & \texttt{cd167d7d34197d98} & 427 & 29871 & \texttt{SPIECE\_UNDERLINE} \\
		2 & \texttt{bec72f2782bf3be5} & 353 & 450   & \texttt{\_The} \\
		3 & \texttt{1dfdd463904c2fcf} & 218 & 13    & \texttt{newline} \\
		4 & \texttt{c057b29aa81c90fb} & 218 & 13    & \texttt{newline} \\
		5 & \texttt{41cbbddc2cb91529} & 124 & 319   & \texttt{\_A} \\
		\hline
	\end{tabular}
\end{table}

In practical deployments, these highly repetitive tokens are prime candidates for KV 
caching and are explicitly mitigated by GELO’s cross-user batch mixing and token-flooding detection
 (Section~\ref{sec:core-algorithm}). After filtering these few repetitive cases, the embeddings are overwhelmingly
  unique and high-entropy, which thwarts simple frequency-analysis and known-plaintext attacks.

\subsubsection{Geometric and Dimensionality Priors}

Even if all embeddings are unique, an attacker can exploit geometric structure in the embedding space. We analyzed 3.5 million filtered
(unique) embeddings and identified two key structural properties.

\paragraph{Distribution of norms}
We measured the $L_2$ norm of each 4096-dimensional embedding. The data exhibits a strong structural prior:
\begin{itemize}
	\item Mean norm: $24.18$
	\item Std. dev. of norm: $0.954$ (coefficient of variation: $0.039$)
\end{itemize}
The extremely low variance indicates that nearly all embedding vectors lie on a hypersphere of radius $\approx 24$. This is informative
non-Gaussian structure: the sources are not uniformly distributed in $\mathbb{R}^{4096}$ but constrained to a narrow shell. Such structured geometry is consistent with prior observations that contextualized language-model representations are anisotropic rather than isotropic Gaussian clouds~\cite{embedd1ethayarajh}.

\paragraph{Effective dimensionality (PCA)}
We performed Principal Component Analysis (PCA) to estimate intrinsic dimensionality:
\begin{itemize}
	\item Full dimension ($d$): $4096$
	\item Participation ratio (PR): $123.16$
\end{itemize}
Thus, the hidden states effectively lie on a low-dimensional manifold, roughly $33\times$ smaller than the ambient space.

\paragraph{Implication for security}
An adversary need not solve a full $4096 \times 4096$ de-mixing problem. 
As discussed in Section~\ref{sec:gram-leaks}, 
the covariance leak ($U^\top U = H^\top H$) reveals the principal subspace. 
The attacker can project $U$ onto this $\sim 123$-dimensional subspace and attempt a much smaller, 
potentially more tractable, $123 \times 123$ BSS problem. This informs the setup of our deobfuscation attack 
in the next section.

\subsubsection{Anchor-Based Recovery Attacks}
\label{sec:anchor-recovery}

We consider a potential known-plaintext (anchor-based) attack: if an adversary knows (or correctly guesses) $k$ of the $n$ tokens in a
batch, can they leverage this information to deobfuscate the remaining $n-k$ unknown tokens?

\paragraph{Attack simulation}
We simulate a best-case scenario for the attacker.
\begin{itemize}
	\item \textbf{Attacker’s knowledge.} The attacker knows $k$ “anchor” rows $H_K$, which are a subset of the true rows of $H$.
\end{itemize}
The attacker’s goal is to use $H_K$ to recover the unknown rows of $H$. We evaluate three attack variants.

\paragraph{Attack methodologies}
The first step is to estimate the mixing rows $A_K$ corresponding to the known anchors $H_K$ via ridge least squares:
\begin{equation}
	A_K \;=\; U H_K^\top \,\big(H_K H_K^\top + \lambda I\big)^{-1} \, ,
\end{equation}
with regularization $\lambda > 0$. Given $A_K$, the attacker proceeds with one of the following:

\begin{enumerate}
	\item \textbf{Deflation (subtraction).} Subtract the anchor contribution and run BSS (e.g., ICA via the FastICA fixed-point solver~\cite{hyvarinen1999fastica,hyvarinen2000ica}) on the residual:
	\begin{equation}
		U_{\mathrm{res}} \;=\; U \;-\; A_K H_K \, .
	\end{equation}
	
	\item \textbf{Projection.} Project onto the subspace orthogonal to the anchor subspace and run BSS on the residual:
	\begin{align}
		P_A &= A_K \big(A_K^\top A_K\big)^{-1} A_K^\top \,, \\
		U_{\mathrm{res}} &= \big(I - P_A\big)\, U \, .
	\end{align}
	This strictly removes anchor leakage but also removes significant signal energy.
	
	\item \textbf{Constrained ICA.} Use the anchor subspace as a hard constraint. Construct an orthogonal basis $B$ that aligns the first $k$
rows with the anchor subspace, then rotate
	\begin{equation}
		U_{\mathrm{rot}} \;=\; B^\top U \;=\; \begin{bmatrix} Z_{\mathrm{top}} \\[2pt] Z_{\perp} \end{bmatrix} \, ,
	\end{equation}
	and run BSS only on $Z_{\perp}$ to recover the unknown signals. In practice, residual ICA is performed in a reduced row subspace ($r
\approx n-k$) to avoid overfitting when anchors are numerous.
\end{enumerate}

A detailed description of the attack pipeline is provided in the appendix.

\paragraph{Results: recovery quality vs. known anchors}
We quantify attack success via the 95th-percentile (p95) cosine similarity between the attacker’s recovered vectors and the true, unknown
hidden states. A value of 1.0 indicates perfect recovery; values near 0.0 indicate failure.

\begin{table}[t]
	\centering
	\caption{Non-anchor recovery quality ( 95th-percentile of cosine similarity) versus number of known anchors $k$.}
	\label{tab:anchor-recovery}
	\begin{tabular}{rccc}
		\hline
		Known anchors ($k$) & Projection & Subtraction  & Constrained ICA \\
		\hline
		0 & 0.341 & 0.341 & 0.341 \\
		2 & 0.323 & 0.323 & 0.328 \\
		5 & 0.317 & 0.315 & 0.316 \\
		10 & 0.301 & 0.310 & 0.258 \\
		20 & 0.283 & 0.288 & 0.237 \\
		40 & 0.245 & 0.256 & 0.207 \\
		100 & 0.207 & 0.231 & 0.185 \\
		200 & 0.279 & 0.276 & 0.277 \\
		240 & 0.408 & 0.370 & 0.375 \\
		\hline
	\end{tabular}
\end{table}

It is visible in Table~\ref{tab:anchor-recovery} that as $k$ increases, the recovery quality for the remaining unknown tokens generally
decreases markedly up to the point when more than 90\% of rows are anchors. This counter-intuitive trend
highlights a key strength of GELO: projection-based defences must remove the anchor subspace, but this also removes signal energy and
distorts the residual, often making it more Gaussian—conditions under which ICA/BSS is less effective.

\subsubsection{Results: Geometric Recovery}

Beyond per-vector similarity, we measure an attacker’s ability to recover the geometric structure of the unknown data. We quantify this
using a matched-subset Gram error, which evaluates how well the pairwise dot products (i.e., geometry) of the recovered tokens match those
of the true tokens.

Metric definition:
\begin{itemize}
	\item \textbf{Matching.} For the $n-k$ unknown true rows in $H$ and the attacker’s corresponding estimates $\hat{H}$, we find an optimal
one-to-one pairing using the Hungarian algorithm~\cite{kuhn1955hungarian} with costs based on absolute cosine similarity.
	\item \textbf{Subsets.} This yields two matched subsets: $H_{\text{sub}}$ (true rows) and $\hat{H}_{\text{sub}}$ (estimated rows).
	\item \textbf{Gram matrices.} Compute row-side Gram matrices
	\begin{equation}
		G_{\text{true}} = H_{\text{sub}} H_{\text{sub}}^\top, 
		\qquad
		G_{\text{est}} = \hat{H}_{\text{sub}} \hat{H}_{\text{sub}}^\top \, .
	\end{equation}
	\item \textbf{Error metric.} The relative Frobenius error is
	\begin{equation}
		\mathrm{GramError} \;=\; \frac{\lVert G_{\text{est}} - G_{\text{true}} \rVert_F}{\lVert G_{\text{true}} \rVert_F} \, .
	\end{equation}
\end{itemize}

A high error ($\gg 1.0$) indicates failure to reconstruct the structural geometry.
 Note that the row-side Gram matrix $G_{\text{true}}$ is not constrained by the feature-side 
 covariance $U^\top U = H^\top H$ leaked under orthogonal $A$ (Section~\ref{sec:gram-leaks}).
  Moreover, the attacker’s residualization (projection or subtraction) alters $U$, 
  breaking such identities and leaving no “free” information about $G_{\text{true}}$.

\begin{table}[t]
	\centering
	\caption{Geometric recovery error (median Gram error; $n=512$).}
	\label{tab:gram-n512}
	\begin{tabular}{rccc}
		\hline
		Known anchors ($k$) & Constrained ICA & Subtraction & Projection \\
		\hline
		0 & 1.414 & 1.414 & 1.414 \\
		2 & 0.805 & 0.852 & 0.802 \\
		5 & 0.806 & 0.802 & 0.807 \\
		10 & 0.840 & 0.777 & 0.840 \\
		20 & 0.871 & 0.811 & 0.872 \\
		40 & 0.888 & 0.861 & 0.889 \\
		100 & 0.896 & 0.892 & 0.897 \\
		200 & 0.898 & 0.899 & 0.899 \\
		240 & 0.898 & 0.900 & 0.900 \\
		\hline
	\end{tabular}
\end{table}

Table~\ref{tab:gram-n512} shows that with $k \approx 10$–20 known anchors, the attacker reduces the relative Gram error from $\sim 1.41$
(no anchors) to $\sim 0.78$–0.87, i.e., a $\sim 40\%$ improvement—substantial, yet still far from accurate reconstruction of non-anchor
correlations.

\begin{table}[t]
	\centering
	\caption{Geometric recovery error under strong Gaussian shielding noise (median Gram error; $n=256$).}
	\label{tab:gram-noise}
	\begin{tabular}{rccc}
		\hline
		Known anchors ($k$) & Constrained ICA & Subtraction & Projection \\
		\hline
		0 & 64.640 & 64.640 & 64.640 \\
		2 & 63.722 & 64.064 & 63.129 \\
		5 & 62.044 & 63.715 & 60.755 \\
		10 & 58.805 & 62.994 & 56.117 \\
		20 & 49.861 & 62.818 & 45.229 \\
		40 & 32.339 & 61.120 & 26.549 \\
		100 &  7.504 & 58.394 &  4.011 \\
		200 &  1.343 & 51.173 &  0.868 \\
		240 &  0.914 & 39.186 &  0.862 \\
		\hline
	\end{tabular}
\end{table}

Table~\ref{tab:gram-noise} setup: $n=256$; strong Gaussian noise via shield rows (5\% extra rows) scaled to 10× the mean row norm; ICA run
with whitening/unwhitening and an $r$-delta to avoid edge cases; metrics aggregated across batches.

Gaussian shield rows provide strong defense against anchor-based attacks: only when anchors exceed roughly 70\% of rows does the Gram error
approach the clean-data result. Subtraction remains unstable in the presence of noise.

\subsubsection{General Blind Source Separation (BSS) Attacks}
\label{sec:bss-attacks}

While Section~\ref{sec:anchor-recovery} analyzed targeted, anchor-based attacks, we now evaluate GELO’s resilience against general-purpose
Blind Source Separation (BSS) algorithms. The goal is to empirically validate the effectiveness of the mitigations proposed in
Section~\ref{sec:gram-leaks}—specifically, Random Vector Padding (Mitigation~2)—and the token-frequency defenses from
Section~\ref{sec:core-algorithm}.

Experimental setup follows Section~\ref{sec:anchor-recovery}. We apply a suite of BSS algorithms (FastICA~\cite{hyvarinen1999fastica},
JADE~\cite{cardoso1993jader}, and joint diagonalization, JD) to $U$ to recover $H$. Simulations span batch sizes $n \in [64, 1024]$ and
padding noise scales $\in [0, 10]$. Success is measured by
cosine similarity between recovered vectors and true hidden states after optimal one-to-one matching.

\paragraph{Results: Efficacy of Vector Padding}

\begin{figure}[!htbp]
	\centering
	\includegraphics[width=0.95\linewidth]{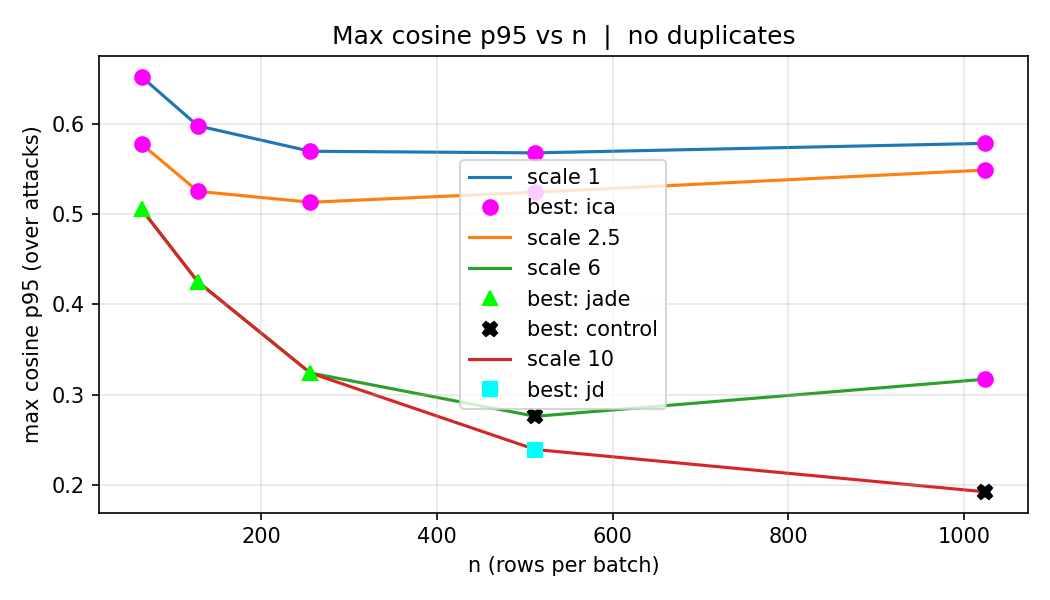}
	\caption{95th percentile cosine similarity of recovered tokens versus padding scale. Each batch is augmented with 5\% random Gaussian
shield rows at varying scales; no exact token repeats are allowed.}
	\label{fig:bss-p95}
\end{figure}

\begin{figure}[!htbp]
	\centering
	\includegraphics[width=0.95\linewidth]{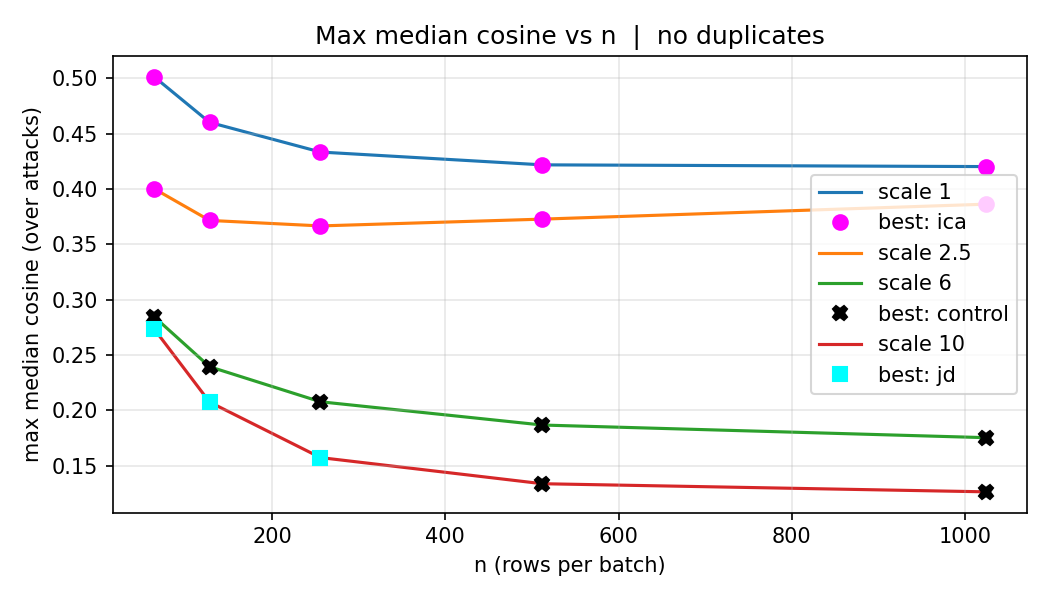}
	\caption{Median cosine similarity of recovered tokens versus padding scale. Each batch is augmented with 5\% random Gaussian shield rows
at varying scales; no exact token repeats are allowed. Marker shapes denote attack method (ICA, JADE, JD).}
	\label{fig:bss-median}
\end{figure}

We report both the median (Figure~\ref{fig:bss-median}) and the 95th percentile (p95) cosine similarity (Figure~\ref{fig:bss-p95}). The
latter reflects the best-quality recoveries within a batch.

The data reveal two key findings:
\begin{enumerate}
	\item \textbf{Raw states are vulnerable.} With no padding (scale $=0$) or low-energy padding (scale $=1$), 
	attacks are partially successful. ICA performs best, achieving median cosine similarity in the $0.42$–$0.53$ range.
	 Tails remain large: at scale $=1$, ICA p95 is high ($\approx 0.97$–$0.99$ for $n \ge 256$), 
	 even when medians are modest. However, as shown in Section~\ref{sec:anchor-recovery}, 
	 recovering a small-to-moderate number of anchors does not translate into recovery of non-anchor rows.
	\item \textbf{Padding mitigation is highly effective.} With high-energy shield vectors (scale $=10$), p95 drops dramatically, remaining
below $0.28$ and reaching as low as $0.13$ for larger, capped batches. Random padding “pollutes” batch statistics, rendering BSS methods
ineffective.
\end{enumerate}

\begin{figure}[!htbp]
	\centering
	\includegraphics[width=0.95\linewidth]{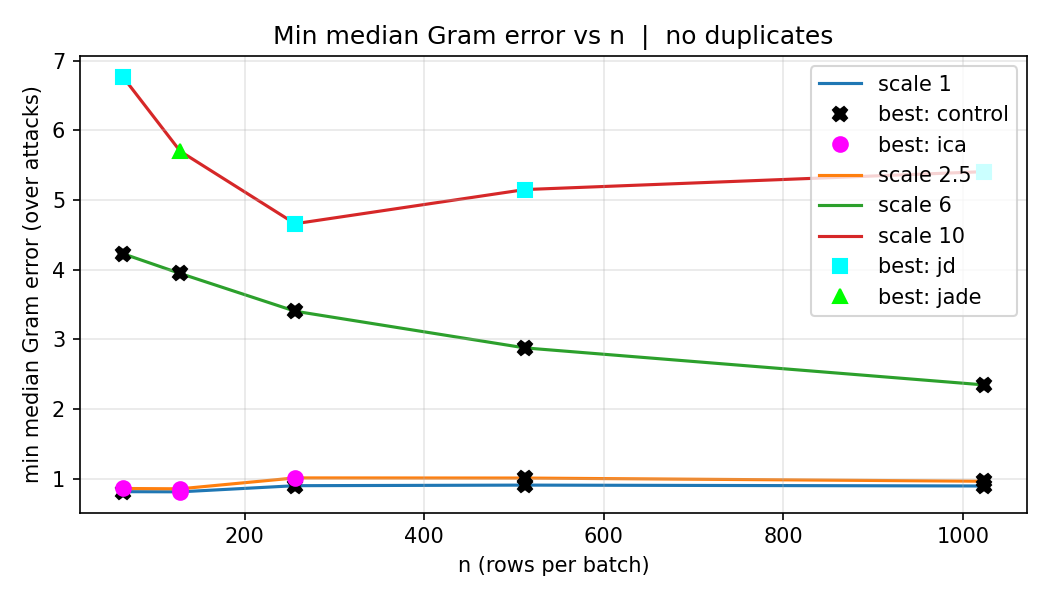}
	\caption{Median Gram error of recovered tokens versus padding scale. Each batch is augmented with 5\% random Gaussian shield rows at
varying scales; no exact token repeats are allowed. Marker shapes denote attack method (ICA, JADE, JD).}
	\label{fig:bss-gram}
\end{figure}

We also evaluate geometric recovery using the matched-subset Gram error (Section~\ref{sec:anchor-recovery}). The results are shown in
Figure~\ref{fig:bss-gram}.
Low-noise regimes show relatively good row-geometry recovery (error $< 1.0$). 
As Gaussian augmentation strengthens,
Gram error inflates at small $n$ but tends to decline with more samples; nevertheless, 
with scale $=10$ it remains high even at $n=1024$ across all methods.

Note: Left mixing with an orthogonal $A$ preserves the feature-side covariance $U^\top U = H^\top H$, but our Gram metric is row-side and
is computed after unmixing/residualization by the attacker, so this identity does not aid geometric recovery.

In summary, these experiments validate GELO’s security model. While raw hidden states can be statistically vulnerable (enabling partial
recovery of frequent tokens), padding each batch with a small number of high-energy random vectors is a practical and highly effective
countermeasure that thwarts general BSS-based deobfuscation attacks.

\subsection{Transformer-Based Recovery Attacks}
\label{sec:learned-attack}

We performed additional experiments with transformer-based attack models to test whether learned sequence models can exploit structure that is
not captured by BSS/ICA methods. For normally distributed hidden states, there are infinitely many pairs $(A,H)$ that produce the same
observation $U = AH$. Intermediate LLM embeddings, however, are not arbitrary Gaussian vectors: embeddings from the same text may contain
correlations that a learned attacker could exploit. We therefore evaluate transformer-based attacks as an empirical stress test of GELO under
structured hidden states.

The attack model is a modified Llama~2-style transformer with positional embeddings and causal masking disabled, so that it operates as a permutation-equivariant encoder over the mixed rows of $U$. 

 The model receives a set of obfuscated activation rows $(U_{\mathrm{obs}}\in\mathbb{R}^{n\times 4096})$ and predicts a set of clean rows $(\hat H\in\mathbb{R}^{n\times 4096})$. We train transformer with 24 layers, hidden size 384, 12 attention heads, head dimension 32, and MLP size 1536. The model has approximately 60M trainable parameters. It is trained with AdamW using learning rate $10^{-5}$, weight decay 0.02, and batch size 16.

\subsubsection{Text-Presence Retrieval Attack}
\label{sec:text-presence-retrieval}
We first evaluate an intentionally easier text-presence retrieval attack. For each training sample, token-level embeddings from several source
texts are combined, optionally augmented with shield rows, and transformed by a fresh random mixing matrix to produce the obfuscated set $U$.
A transformer encoder receives $U$ and uses a learned prepended query token (CLS-style) to produce a single normalized query vector. This
query is matched against a database of precomputed text-level candidate embeddings. The model is trained with multi-positive InfoNCE so that
all true source texts for the sample are ranked above negatives. We report results with both Llama~2~\cite{llama2} and a Qwen-family 9B
candidate-embedding model~\cite{qwen3}. Evaluation reports positive probability mass on the training candidate pool and on held-out
validation mixtures.

\begin{table}[!htbp]
	\centering
	\caption{Transformer-based learned retrieval attack results across shield-vector distributions, shield-fractions, and scales. Each result reports final retrieval mass as \texttt{train} / \texttt{val}, where \texttt{train} is measured on the training candidate pool and \texttt{val} is measured on the held-out validation split. Chance positive mass = 0.0625 }
	\label{tab:learned-retrieval-attack}
		\setlength{\tabcolsep}{3pt}
		\scriptsize
		\begin{tabular}{llllcc}
			\hline
			Kind & Fraction range & Scale range & Randomization & llama2 train/val & Qwen train/val \\
			\hline
			Manifold & 0.05 & 1 & none & 0.8318 / 0.8212 & 0.8333 / 0.7561 \\
			Manifold & 0.01--0.14 & 5--30 & log-uniform & 0.0704 / 0.0723 & 0.5370 / 0.5401 \\
			Manifold & 0.05 & 5 & none & 0.5330 / 0.4990 & 0.5845 / 0.5214 \\
			Gaussian & 0.05 & 2.5--20 & log-uniform & 0.4584 / 0.3700 & 0.4890 / 0.4094 \\
			Gaussian & 0.05 & 10 & none & 0.4749 / 0.5010 & 0.4554 / 0.3907 \\
			Student-$t$ & 0.05 & 10 & none & 0.3115 / 0.3042 & 0.3659 / 0.3828 \\
			Manifold & 0.05 & 2.5--20 & log-uniform & 0.3641 / 0.3737 & 0.4227 / 0.3509 \\
			Gaussian & 0.01--0.09 & 10 & log-uniform & 0.3949 / 0.3964 & 0.3741 / 0.3356 \\
			Gaussian & 0.03--0.09 & 2.5--20 & log-uniform & 0.4291 / 0.3557 & 0.4143 / 0.3232 \\
			Gaussian & 0.01--0.12 & 2.5--25 & log-uniform & 0.3236 / 0.3316 & 0.2892 / 0.2975 \\
			Manifold & 0.05 & 10 & none & 0.2238 / 0.2251 & 0.2741 / 0.2634 \\
			Gaussian & 0.05 & 5 & none & 0.7190 / 0.6952 & 0.6774 / 0.6954 \\
			\hline
		\end{tabular}
\end{table}

Overall, the results in Table~\ref{tab:learned-retrieval-attack} show that learned transformer-based retrieval is sensitive to the strength
of the shielding distribution. Weak manifold-aligned shielding remains partially vulnerable, but increasing the shield scale substantially
lowers validation positive mass. This supports the use of high-energy shield vectors as a practical mitigation against aggregate text-level
leakage.

This retrieval task is intentionally favourable to the attacker: the model is not required to reconstruct individual hidden-state rows, but only to detect which source texts are present in the mixed batch. Thus, the experiment measures whether any aggregate text-level signal remains
detectable in $U$, rather than whether the attacker can invert the mixing or recover token embeddings.

\subsubsection{Token-Level Unmixing Attack}
\label{sec:token-level-unmixing}
We evaluate a stricter token-level unmixing attack, where the model is trained to map the obfuscated batch $U$ directly back to source embeddings $H_i$. This task is substantially harder because it requires resolving the unknown per-batch mixing transform rather than merely detecting text presence. In our experiments, transformer-based token recovery fails under strong fresh mixing, providing further evidence that residual aggregate signals do not translate into practical reconstruction of hidden states.

• Unmix Attack Setup Overview
We evaluate deobfuscation with a learned unmixing attack that maps mixed token sets back to latent clean embeddings. Each training sample is built by sampling token rows from multiple texts, adding optional shield rows, and applying a random mixing matrix (A) to obtain observed rows (U). A set-transformer encoder-decoder processes (U)
and predicts a set of candidate recovered rows $\hat{H}$, together with a learned linear unmix operator $\hat{A}$.

Training uses two coupled objectives:

1. Cycle-consistency loss

$\mathcal{L}_{cycle}=|\hat{A}\hat{R}-U|_2^2$

which enforces that remixed predictions reconstruct observed mixtures.

2. Set matching loss via Sinkhorn optimal transport~\cite{cuturi2013sinkhorn}

A Sinkhorn-based optimal transport plan aligns $\hat{H}$ to target clean rows H under squared Euclidean cost, providing permutation-invariant supervision:

$\mathcal{L}_{OT}=\langle P_{sinkhorn}, C(\hat{H},H)\rangle$

Additional regularization terms stabilize training (mixing-matrix regularization $||\hat A||_F^2$) and row-norm regularization, and optional dustbin/unbalanced transport variants handle unmatched mass in partial-observation settings.

Row norm regularisation is squared difference between average norm of predicted rows $\hat H$ and target rows $H$.

\[
\mathcal{L}_{\mathrm{row}}
= \left(\frac{1}{\hat{n}} \sum_{i=1}^{\hat{n}} \lVert \hat{H}_i \rVert_2
- \frac{1}{n} \sum_{j=1}^{n} \lVert H_j \rVert_2\right)^2 .
\]

To monitor attack success, we compute a global-assignment same-embedding probability mass. Predicted rows are first matched to a candidate pool of true embedding rows using Hungarian assignment. We then measure how much probability mass is assigned to the correct target embedding for each prediction. The \texttt{train} value is computed against the training candidate pool, while \texttt{val} is computed on held-out validation set. Higher values indicate more successful recovery; chance-level mass is approximately 0.065.

\begin{table}[!htbp]
	\centering

	\caption{Token-level unmixing attack results. The final column reports global-assignment same-embedding probability mass as \texttt{train} / \texttt{val}, where \texttt{train} is measured against the training candidate pool and \texttt{val} on held-out validation set. Higher is better for the attacker; chance mass is approximately 0.065.}

	\label{tab:unmix-attack}
	\resizebox{\linewidth}{!}{%
        \begin{tabular}{lllllll}
			\hline
			Shield & Positives/text & Tokens/text & Obs rows & Sinkhorn ($\epsilon$, iters) & OT weight & train/val \\
			\hline
			manifold, frac 0.01--0.10, scale 1--20 & 2--8 & 8--96 & 0.96 & (5.0, 50) & 0.1 & 0.0788 / 0.0678 \\
			manifold, frac 0.01--0.10, scale 1--20 & 2--8 & 8--96 & 0.80 & (5.0, 50) & 1 & 0.0750 / 0.0691 \\
			gaussian, frac 0.05, scale 5 & 2--8 & 8--96 & 0.96 & (5.0, 50) & 0.1 & 0.0703 / 0.0625 \\
			gaussian, frac 0.05, scale 5 & 4 & 64 & 0.96 & (5.0, 50) & 0.1 & 0.2800 / 0.2712 \\
			gaussian, frac 0.05, scale 5 & 4 & 64 & 0.96 & (0.1, 30) & 1 & 0.4181 / 0.4144 \\
			gaussian, frac 0.00, scale 5 & 4 & 64 & 1 & (5.0, 50) & 0.1 & 0.4431 / 0.4375 \\
			none (gauss frac 0.00) & 32--64 & 4--8 & 0.96 & (5.0, 50) & 0.1 & 0.0725 / 0.0703 \\
			none (gauss frac 0.00) & 32--64 & 4--8 & 0.96 & (5.0, 50) & 1 & 0.0712 / 0.0650 \\
			none (gauss frac 0.00) & 16--32 & 8--16 & 0.96 & (5.0, 50) & 0.1 & 0.1150 / 0.1144 \\
			none (gauss frac 0.00) & 8--16 & 16--32 & 0.96 & (5.0, 50) & 0.1 & 0.1494 / 0.1544 \\
			none (gauss frac 0.00) & 4--8 & 32--64 & 0.96 & (5.0, 50) & 0.1 & 0.2062 / 0.2216 \\
			\hline
		\end{tabular}%
	}
\end{table}

Table~\ref{tab:unmix-attack} shows that strong mixing and shielding drive validation mass close to chance, indicating failed token-level embedding recovery. In contrast, lighter regimes without shielding or with easier candidate structure yield higher same-embedding mass. Test runs without shielding demonstrate that main defence comes from mixing more texts.   This behavior suggests that repeated exposure to multiple embeddings from the same source text provides the attacker with exploitable in-text co-variation: rows from the same text are not independent samples, but share contextual and semantic structure. Mixing more independent texts dilutes this correlation signal, making it harder for the learned model to associate rows and infer the underlying clean embeddings.

\section{Security Analysis and Identifiability}
\label{sec:security}

\subsection{Why GELO works: non-identifiability by design}

\paragraph{Algebraic ambiguity (GL($n$) invariance).}
For a batch of $n$ token vectors, the accelerator observes the mixed hidden states $U = A H$, where $A \in \mathbb{R}^{n \times n}$ is
invertible and $H \in \mathbb{R}^{n \times d}$. The observation $U$ constrains only the product $A H$. For any invertible
$R \in \mathrm{GL}(n)$, the pairs $(A R^{-1}, R H)$ yield the same $U$. Hence $H$ is identifiable only up to an unknown invertible
transform. Without side information that ties $R$ to the true token basis, the attacker cannot uniquely recover $H$ from $U$ alone.

\paragraph{Dynamic mixing prevents accumulation.}
If $A$ is freshly and independently sampled for every batch, then statistics from different batches do not align in a common coordinate
system. The attacker cannot “average out” the mixing to estimate a stable inverse, in stark contrast to static obfuscation (e.g., fixed
permutations), which is vulnerable to multi-run statistical attacks.

\paragraph{BSS/ICA and learned-attack assumptions are weakened.}
ICA exploits non-Gaussianity and independence of sources under a fixed mixing matrix. Modern hidden states are high-dimensional,
correlated, and structured; moreover, we refresh $A$ per batch, invalidating the fixed-mixing assumption. As a result, off-the-shelf ICA
and dictionary-learning methods lack the stationary signal they need to converge.

Learned attackers are more flexible: a transformer-based model can exploit distributional and within-text correlations that are not captured
by classical BSS assumptions. Our experiments therefore separate aggregate leakage from reconstruction. Text-presence retrieval is an easier
stress test and can detect residual text-level signal under weak shielding, but this does not remove the algebraic ambiguity of $U=AH$ or imply
row-level recovery of $H$. In the stricter token-level unmixing experiment, learned recovery fails under strong fresh mixing and shielding.

\subsection{Information-theoretic view}

\paragraph{No cross-batch gain under fresh mixing.}
Index batches by $t$, with $U_t = A_t H_t$ and $A_t$ independent across $t$ and independent of $H_t$. Then for any fixed batch $t$, the
other obfuscated batches provide no extra information about $H_t$ beyond $U_t$:
\[
I(H_t; U_{1:T}) \;=\; I(H_t; U_t).
\]
Intuition: $U_i$ for $i \neq t$ depend on independent nuisance variables $A_i$ and independent hidden states $H_i$, so they are
conditionally irrelevant for $H_t$ once $U_t$ is known.

\paragraph{Second-order statistics do not resolve $H$.}
Across samples within a batch, the observable second-order structure factors as
\[
\Sigma_U \;=\; A \, \Sigma_H \, A^\top.
\]
Here $\Sigma_H$ is the covariance of the hidden-state features across tokens in a batch. Without $A$, whitening reduces the problem only to
an unknown orthogonal (or more generally, invertible) ambiguity; higher-order statistics that ICA would use require a fixed mixing and
favorable source assumptions, which we intentionally avoid.

\subsection{How many anchors would suffice?}

Consider an attacker who (unrealistically) knows the full hidden-state matrix $H$ for a batch and can observe the corresponding mixed
matrix $U = A H$. Then the mixing matrix is algebraically recoverable as
\[
A \;=\; U\, H^{+},
\]
and the batch could be fully de-mixed. Note that $H \in \mathbb{R}^{n \times d}$ is typically non-square, with $d \gg n$ in LLMs.
However, learning $H$ in full is precisely the privacy breach we aim to prevent. With partial
in-batch side information (e.g., a small number of ``anchor'' token vectors), the problem remains underdetermined because the left-mixing
matrix $A \in \mathbb{R}^{n \times n}$ couples \emph{all} token rows. Our empirical results in
Section~\ref{sec:anchor-recovery} show that such partial anchors do not enable recovery of the remaining tokens under our mitigations.

\subsection{Batch accumulation: upper limits and empirical evidence}

\paragraph{Fresh $A$ per batch.}
There is no principled benefit from storing many $U$’s. Each batch comes with its own unknown transform, so cross-batch alignment is
impossible without side information. Accumulation cannot reduce the core ambiguity beyond invariants that survive unknown invertible
transforms (e.g., rank).

\paragraph{Fixed $A$ (not our setting).}
If $A$ were fixed and the sources satisfied ICA’s identifiability assumptions (independent, suitably non-Gaussian, at most one Gaussian),
then with many samples one could estimate $A$ up to permutation and scaling. GELO’s design specifically avoids this setting by refreshing
$A$ each batch.

\paragraph{Empirical observation.}
Running multiview ICA~\cite{NEURIPS2020_de03beff} (implemented via picard-ICA) on many batches did not improve reconstruction compared to a
single batch (around 0.2 cosine similarity on 10 batches). This is consistent with the theory above: without a fixed mixing, cross-batch
statistics do not concentrate toward an invertible unmixing.

\subsection{Formal statements we can claim}

\begin{itemize}
	\item Per-batch non-identifiability: $U = A H$ reveals $H$ only up to an unknown invertible transform; without side information, $H$ is
	not uniquely recoverable from a single batch.
	\item No cross-batch gain with fresh mixing: under independence of $A_t$ across batches, other batches do not increase information about
	$H_t$ beyond $U_t$.
	\item Empirical validation against classical attacks: our experiments show that off-the-shelf and constrained ICA variants fail to recover
usable hidden states; storing many batches does not help in practice.
	\item Empirical validation against learned attacks: transformer-based retrieval can detect aggregate text-level signal in weakly shielded
mixtures, but transformer-based token-level unmixing fails under strong fresh mixing and shielding.
\end{itemize}

\paragraph{Not a cryptographic proof.}
We do not offer a reduction-based or complexity-theoretic proof of security. Our argument is an identifiability analysis under a stated
threat model, supported by negative empirical results against both classical and learned recovery attacks. This is appropriate for obfuscation
(as opposed to encryption) and aligned with GELO’s security model based on dynamic per-batch mixing, limited side information, and empirical
failure of practical deobfuscation methods.

\subsection{Practical implications}

\begin{itemize}
	\item To remain secure against accumulation, the mixing must be refreshed per batch.
	\item Avoid predictable structure that could act as anchors; any auxiliary side information that ties $U$ back to $H$ across batches weakens security.
	\item If an attacker obtains enough \emph{in-batch} side information to solve for the mixing matrix $A$ (e.g., many exact anchors with correct correspondences), then the remaining tokens in that batch can be de-mixed algebraically. System design should make such leakage implausible.
	\item Learned text-presence signals should be treated as leakage indicators, not as reconstruction evidence. Stronger shielding and direct
		token-level recovery tests are necessary to distinguish aggregate detection from practical hidden-state recovery.
\end{itemize}

Taken together, these results explain why GELO resists deobfuscation: the attacker’s problem is under-determined by design, cross-batch
aggregation offers no principled advantage under fresh mixing, and practical recovery methods fail to produce meaningful token-level
reconstructions under strong mixing and shielding. The learned text-presence results make the analysis more conservative: they show that
aggregate distributional leakage can exist in weak regimes, while the harder unmixing experiments indicate that such leakage does not translate
into practical recovery of hidden states.

%
%
\section{Conclusion and Future Work}
\label{sec:conclusion}
We presented GELO, a ``good-enough'' privacy layer for LLM inference on untrusted accelerators. GELO keeps sensitive activations inside a
TEE while offloading the dominant linear projections in attention (notably the $Q/K/V$ GEMMs; and $O$ in our functional-equality test). The
TEE applies fresh, per-batch
left mixing $U = A H$ before offload and unmixes on return, guaranteeing exact correctness in exact arithmetic and near-identical outputs in
low precision. In our prototype on Llama~2 7B, GELO preserves functional behavior (e.g., $\ge$98.8\% top-1 token equality in bfloat16).
Controlled microbenchmarks show about $20$--$30\%$ compute-side overhead near favorable batch sizes, whereas the unoptimized asynchronous
remote prototype has much higher end-to-end overhead because it is dominated by Python/socket serialization, RPC scheduling, and activation
transfer rather than mixing or GEMM.
By never reusing $A$, GELO prevents cross-batch statistical accumulation and reduces deobfuscation to a single-batch BSS problem. Our analysis
identifies key leakage channels (e.g., Gram-matrix invariants under orthogonal $A$) and shows that non-orthogonal mixing or a small fraction
of high-energy shield vectors can effectively mitigate practical ICA/BSS and anchor-based attacks at modest overhead.

There are several promising directions for future work:
\begin{itemize}
	\item \textbf{Integration with LLM engines and KV caching.} Integrate GELO into inference engines such as vLLM, especially around
	KV-cache management with the key goal of maintaining high throughput without compromising users' data due to cache-sharing leaks
	\cite{iknowwhatyouask}.

	\item \textbf{Stronger formalization.} Strengthen the identifiability analysis into tighter theorems under explicit assumptions on hidden
	state priors and attacker observations, and quantify leakage from invariants (e.g., Gram matrices) under practical mitigations.

	\item \textbf{Broader coverage and quantization.} Extend selective offload beyond $Q/K/V/O$ (e.g., MLP projections) and characterize
	numerical stability under common inference precisions and quantization (BF16/FP16/FP8/INT8).

	\item \textbf{Reducing overhead in large batches.} Explore faster constructions for fresh, well-conditioned mixing (e.g., structured
	orthogonal transforms) and system optimizations that reduce the communication bottleneck observed in our prototype.
	\item \textbf{Stronger adversaries and side channels.} Evaluate adaptive prompt-selection attackers under realistic rate limits and
	batching policies, and extend the threat model to include accelerator-side side channels (timing, traffic patterns, and cache effects).
\end{itemize}

\section*{Acknowledgements}
\noindent \textbf{Funding:} This work was supported by SingularityNET Foundation.

\section*{Declaration of Generative AI and AI-assisted technologies in the manuscript preparation process}
\noindent During the preparation of this manuscript, the authors used OpenAI GPT-5.1-5.5 to assist with drafting and refactoring code for attack
baselines and to help improve clarity of the article. All AI-assisted outputs were reviewed, tested where applicable, and edited by the
authors, who take full responsibility for the content of the manuscript.

\appendix
\label{app1}
\section{Computational Complexity}
\label{sec:complexity}

For the Llama~2 7B model with hidden dimension $d=4096$, linear projections account for a significant portion of per-token computation,
especially for typical prompt sizes. Below is a breakdown of multiply-adds (MAdds) for a non-cached token in a decoder layer.

\paragraph{Linear projections}
\[
\text{Q/K/V/O: } \quad 4 \cdot 4096 \times 4096 \;\Rightarrow\; \approx 67\ \text{M per token.}
\]

\paragraph{Feed-forward network (gate, up, down)}
\[
3 \cdot 4096 \times 11008 \;\Rightarrow\; \approx 135\ \text{M per token.}
\]

\paragraph{Core attention} Using multi-head attention with $h=32$ heads and $d_{\text{head}} = d/h = 128$:
\[
\text{Scores: } Q K^\top \in \mathbb{R}^{1 \times L} \quad \Rightarrow \quad \approx 4096 \cdot L \ \text{MAdds per token},
\]
\[
\text{Weighted sum: } \text{Scores} \cdot V \quad \Rightarrow \quad 32 \cdot (1 \times L)(L \times 128) \approx 4096 \cdot L \ \text{MAdds
per token}.
\]
Thus, the attention core scales as $\approx 8192 \cdot L$ MAdds per token, where $L$ is the total sequence length (cached context plus the
new token).

\paragraph{Crossover point}
Comparing attention ($\sim 8192 L$) to projections+FFN ($\sim 202$M):
\[
8192\,L \;\approx\; 202 \times 10^{6} \quad \Rightarrow \quad L \approx 24{,}658.
\]
For Llama~3-70B with $d=8192$, the crossover occurs at roughly $L \approx 49{,}000$. Consequently, linear projections are the primary
target for offloading in practical prompt lengths.

\section{Anchor-Based Attack}
\label{sec:anchor-attack}

This appendix describes the single end-to-end anchor-based attack pipeline used in
Section~\ref{sec:anchor-recovery}. Starting from a batch of obfuscated hidden states and a small
set of known ``anchor'' tokens, the attacker attempts to reconstruct the remaining hidden states.

\paragraph{Preprocessing.}
\begin{enumerate}
	\item Batch centering: subtract per-feature means from $H$ before mixing.
	\item Optional Gaussian shielding: augment $H$ with random Gaussian rows, $X = [\,H;\ G\,]$, as a defense.
\end{enumerate}

\paragraph{Anchor selection and ridge estimation.}
Choose $k$ anchor rows $H_K$ (oracle-known or detected). The observed obfuscated data is $U = A H$. Estimate the corresponding mixing rows
$A_K \in \mathbb{R}^{n \times k}$ by ridge least squares:
\begin{align}
	G &= H_K H_K^\top, \\
	A_K &= (U H_K^\top)\, \big(G + \lambda_{\mathrm{reg}} I\big)^{-1}.
\end{align}
Here $\lambda_{\mathrm{reg}}>0$ stabilizes inversion when anchors are correlated or ill-conditioned.

\paragraph{Residual construction.}
Given $A_K$, form residuals in one of three ways:
\begin{enumerate}
	\item Subtraction:
	\begin{equation}
		U_{\mathrm{res}} \;=\; U \;-\; A_K H_K.
	\end{equation}
	\item Projection:
	\begin{align}
		P_A &= A_K \big(A_K^\top A_K + \lambda I\big)^{-1} A_K^\top, \\
		U_{\mathrm{res}} &= (I - P_A)\, U.
	\end{align}
	\item Constrained ICA (row-space separation):
	\begin{align}
		&\text{Construct an orthonormal basis } B \text{ that spans anchors and their orthogonal complement,} \\
		&U_{\mathrm{rot}} \;=\; B^\top U \;=\; \begin{bmatrix} Q \\ Q_\perp \end{bmatrix} U,
	\end{align}
	where $B = [\,Q\ \ Q_\perp\,]$ is the orthonormal basis constructed in the next paragraph.
	Then take the last $k_{\mathrm{eff}}$ rows of $U_{\mathrm{rot}}$ (the orthogonal complement) to form
	$U_{\mathrm{res}}$.
\end{enumerate}

\paragraph{Constructing the constrained basis.}
\begin{enumerate}
	\item Orthonormalize $A_K$ to get the anchor-span basis $Q$:
	\[
	Q = \mathrm{qr}(A_K)\ \ (\text{$n \times k$ with orthonormal columns}).
	\]
	\item Compute effective rank $k_{\mathrm{eff}} = \mathrm{rank}(A_K)$.
	\item Form the orthogonal complement by projecting a random matrix $R \in \mathbb{R}^{n \times (n-k_{\mathrm{eff}})}$:
	\[
	Q_{\mathrm{perp}} = (I - Q Q^\top)\, R,\quad Q_\perp = \mathrm{qr}(Q_{\mathrm{perp}}).
	\]

	Here $Q$ spans the anchor subspace and $Q_\perp$ spans its orthogonal complement.
	
	\item Let $U = [\,Q\ \ Q_\perp\,]$ (orthonormal).
\end{enumerate}

\paragraph{Whitening and dimensionality reduction.}
Choose the number of independent row directions $r$ to feed into ICA/FOBI after whitening (trade-off between coverage and stability),
typically $r = n - k - 1$.
\begin{enumerate}
	\item ZCA (symmetric) whitening: compute
	\[
	\frac{U_{\mathrm{res}} U_{\mathrm{res}}^\top}{d} = V_\Lambda \, \Lambda \, V_\Lambda^\top,\quad
	W = V_\Lambda\, \Lambda^{-1/2}\, V_\Lambda^\top,\quad
	Z = W\, U_{\mathrm{res}}.
	\]
	\item PCA clipping: let $U_r \in \mathbb{R}^{n \times r}$ be the matrix whose columns are the $r$
	largest-variance eigenvectors from $V_\Lambda$ (i.e., its first $r$ columns), and project onto this
	row subspace:
	\[
	Z_r = U_r^\top Z.
	\]
\end{enumerate}

\paragraph{BSS and back-projection.}
Run FOBI and/or ICA (e.g., FastICA~\cite{hyvarinen1999fastica}) on $Z_r$. The solver returns an unmixing matrix
$W_r \in \mathbb{R}^{r \times r}$ (orthogonal in the whitened $r$-space) and corresponding estimated sources/components
$S_r \in \mathbb{R}^{r \times d}$ such that
\[
S_r \;\approx\; W_r Z_r,
\]
i.e., each row of $S_r$ is one recovered independent component in the reduced row subspace. Lift back:
\begin{align}
	R_{\mathrm{full}} &= U_r\, W_r\, U_r^\top \;+\; (I - U_r U_r^\top), \\
	W^{-1} &= \sqrt{d}\, V_\Lambda\, \Lambda^{1/2}\, V_\Lambda^\top, \\
	A &= W^{-1} R_{\mathrm{full}}^\top, \\
	\hat{X} &= A^\top U.
\end{align}

\paragraph{Matching and metrics (non-anchors only).}
\begin{enumerate}
	\item Identify true non-anchors; perform one-to-one matching between recovered and true rows using absolute cosine similarity (Hungarian
assignment). Flip signs on matches to resolve ICA sign ambiguity.
	\item Report:
	\begin{itemize}
		\item p95 cosine on non-anchors: 95th percentile of matched absolute cosines.
		\item Gram error (relative Frobenius) on matched non-anchors:
		\[
		\frac{\lVert H H^\top - \hat{H}\hat{H}^\top \rVert_F}{\lVert H H^\top \rVert_F}.
		\]
	\end{itemize}
\end{enumerate}


.




\end{document}